\documentclass[11pt]{article}
\usepackage{amssymb,amsmath,amsfonts}
\usepackage{graphicx}
\usepackage{graphics}
\usepackage{eepic,epsfig}

\@addtoreset{equation}{section}

\makeatother

\begin{document}

\title{Electromagnetic vacuum fluctuations around \\
a cosmic string in de Sitter spacetime}
\author{A. A. Saharian$^{1}$,\thinspace\ V. F. Manukyan$^{2}$,\thinspace\ N.
A. Saharyan$^{1}$ \\
\\
\textit{$^1$Department of Physics, Yerevan State University,}\\
\textit{1 Alex Manoogian Street, 0025 Yerevan, Armenia}\vspace{0.3cm}\\
\textit{$^{2}$Department of Physics and Mathematics, Gyumri State
Pedagogical Institute,}\\
\textit{4 Paruyr Sevak Street, 3126 Gyumri, Armenia}}
\maketitle

\begin{abstract}
The electromagnetic field correlators are evaluated around a cosmic string
in background of $(D+1)$-dimensional dS spacetime assuming that the field is
prepared in the Bunch-Davies vacuum state. The correlators are presented in
the decomposed form where the string-induced topological parts are
explicitly extracted. With this decomposition, the renormalization of the
local vacuum expectation values (VEVs) in the coincidence limit is reduced
to the one for dS spacetime in the absence of the cosmic string. The VEVs of
the squared electric and magnetic fields, and of the vacuum energy density
are investigated. Near the string they are dominated by the topological
contributions and the effects induced by the background gravitational field
are small. In this region, the leading terms in the topological
contributions are obtained from the corresponding VEVs for a string on the
Minkowski bulk multiplying by the conformal factor. At distances from the
string larger than the curvature radius of the background geometry, the pure
dS parts in the VEVs dominate. In this region, for spatial dimensions $D>3$,
the influence of the gravitational field on the topological contributions is
crucial and the corresponding behavior is essentially different from that
for a cosmic string on the Minkowski bulk. There are well-motivated
inflationary models which produce cosmic strings. We argue that, as a
consequence of the quantum-to-classical transition of super-Hubble
electromagnetic fluctuations during inflation, in the postinflationary era
these strings will be surrounded by large scale stochastic magnetic fields.
These fields could be among the distinctive features of the cosmic strings
produced during the inflation and also of the corresponding inflationary
models.
\end{abstract}

\bigskip


\bigskip

\section{Introduction}

The properties of the quantum vacuum are sensitive to both the local and
global geometrical characteristics of the background spacetime. In this
paper we investigate the electromagnetic vacuum polarization sourced by the
gravitational field and by the nontrivial topology due to the presence of a
straight cosmic string. In order to have an exactly solvable problem, as the
background geometry we will consider a spacetime that is maximally symmetric
in the absence of the cosmic string, namely de Sitter (dS) spacetime. For
the cosmic string a simplified model will be taken in which the local
geometry outside the core is not changed by the presence of the string: the
only effect is the planar angle deficit depending on the mass density of the
string.

In addition to the high degree of symmetry, our choice of dS spacetime as
the background geometry is motivated by its importance in modern cosmology.
In most inflationary models the early expansion of the universe is
approximated by the dS phase (for reviews see \cite{Lind94}). The presence
of this phase before the radiation dominated era naturally solves several
problems in the standard cosmological model. More recently, the astronomical
observations of high redshift supernovae, galaxy clusters, and cosmic
microwave background \cite{Ries98} indicate that at the present epoch the
universe is accelerating and the corresponding expansion is dominated by a
source of the cosmological constant type. In this case, the dS spacetime
appears as the future attractor for the geometry of the universe.
Consequently, the investigation of physical effects in dS spacetime is
important for understanding both the early Universe and its future
evolution. A topic which has received increasing attention is related to
string-theoretical models of dS spacetime and inflation. Recently several
constructions of metastable dS vacua within the framework of string theory
are discussed (see, for instance, reviews \cite{Doug07,Cher15}).

The investigation of the quantum field theoretical effects in dS spacetime
is of considerable interest. These effects may have important implications
in cosmology. During an inflationary epoch, quantum fluctuations in the
inflaton field introduce inhomogeneities which play a central role in the
generation of cosmic structures from inflation. The inflation also provides
an attractive mechanism of producing long-wavelength electromagnetic
fluctuations, originating from subhorizon-sized quantum fluctuations of the
electromagnetic field stretched by the dS phase to superhorizon scales.
After these long-wavelength fluctuations have re-entered the horizon in the
post-inflationary era, they can serve as seeds for cosmological magnetic
fields. Related to this inflationary mechanism for the generation of the
seeds, the cosmological dynamics of the electromagnetic field quantum
fluctuations have been discussed in large number of papers (for reviews see
\cite{Kron94}).

In the present paper we investigate the influence of a cosmic string on the
vacuum fluctuations of the electromagnetic field in background of dS
spacetime (for the effects of inflation on the cosmic strings see, for
instance, \cite{Basu94}). Though the cosmic strings produced in phase
transitions before or during early stages of inflation are diluted by the
expansion to at most one per Hubble radius, the formation of defects near or
at the end of inflation can be triggered by several mechanisms (see \cite%
{Hind11} for possible distinctive signals from such models). They include a
coupling of the symmetry breaking field to the inflaton field or to the
curvature of the background spacetime. Moreover, one can have various
inflationary stages, with linear defects being formed in between them \cite%
{Vile97}. Depending on the underlying microscopic model, there exist several
kinds of cosmic strings. They can be either nontrivial field configurations
or more fundamental objects in superstring theories. The cosmic strings are
among the most popular topological defects formed by the symmetry breaking
phase transitions in the early universe within the framework of the Kibble
mechanism \cite{Vile94}. They are sources of a number of interesting
physical effects that include the generation of gravitational waves,
high-energy cosmic rays, and gamma ray bursts. Among the other signatures
are the gravitational lensing and the creation of small non-Gaussianities in
the cosmic microwave background. The cosmic superstrings, which are
fundamental quantum strings stretched to cosmological scales, were first
considered in \cite{Witt85}. More recently, a mechanism for the generation
of this type of objects with low values of the string tensions is proposed
within the framework of brane inflationary models \cite{Cher15,Hind11,Cope11}%
. In these models the accelerated expansion of the universe is a consequence
of the motion of branes in warped and compact extra dimensions.

Although the specific properties of cosmic strings are model-dependent, they
produce similar gravitational effects. In the simplified model with the
string induced planar angle deficit, the nontrivial spatial topology results
in the distortion of the vacuum fluctuations spectrum of quantized fields
and induces shifts in vacuum expectation values (VEVs) of physical
characteristics of the vacuum state such as the field squared and the
energy-momentum tensor. Explicit calculations of this effect have been done
for scalar, fermion and vector fields (see references given in \cite{Bell14}%
). For charged fields, another important characteristic of the vacuum state
is the VEV of the current density (see \cite{Beze15} for a recent discussion
and references therein). The analysis of the vacuum polarization effects
induced by a cosmic string in dS spacetime for massive scalar and fermionic
fields has been presented in \cite{Beze09,Beze10}. Here we will be concerned
with the combined effects of the background gravitational field and of a
cosmic string on the correlators for the electric and magnetic fields and on
the VEVs of the energy density and squared electric and magnetic fields. The
problem will be considered on the bulk of dS spacetime with an arbitrary
number of spatial dimensions $D$. This is motivated by several reasons. In
discussions of cosmic superstrings, depending on the compactification scheme
of extra dimensions, one can have $3\leqslant D\leqslant 9$. In particular,
this is the case for superstrings formed at the end of brane inflation. The
consideration of electrodynamics in spatial dimensions $D>3$ is a natural
way to break the conformal invariance of the $D=3$ theory. The breaking of
conformal invariance is required in inflationary models for the generation
of large scale magnetic fields. Usually this is done by adding additional
couplings of the electromagnetic field (for example, to the inflaton field)
\cite{Kron94}. A mechanism for the generation of cosmological magnetic
fields, based on the dynamics of electromagnetic fluctuations in models with
$D>3$, has been discussed in \cite{Giov00}. The consideration of quantum
field theories in spatial dimensions other than 3 is also required in
dimensional regularization procedure for the ultraviolet divergences.

The paper is organized as follows. In the next section the background
geometry is described and a complete set of mode functions for the
electromagnetic field is given. In section \ref{sec:2point}, two-point
functions for the vector potential and for the electric field strength are
investigated. The VEV of the electric field squared is discussed in section %
\ref{sec:E2}. The part induced by the nontrivial topology of the cosmic
string is explicitly separated and its asymptotic behavior in various
limiting regions is investigated. The two-point functions corresponding to
the Lagrangian density and the magnetic field are considered in section \ref%
{sec:LM}. The topological contributions in the VEVs of the squared magnetic
field and of the vacuum energy density are investigated. The main results
are summarized in section \ref{sec:Conc}. In Appendix we present the main
steps for the evaluation of the integrals appearing in the expressions for
the two-point functions.

\section{Cylindrical electromagnetic modes}

\label{sec:Modes}

We consider $(D+1)$-dimensional locally dS background geometry described in
cylindrical spatial coordinates $(r,\phi ,\mathbf{z})$, $\mathbf{z=}\left(
z^{3},...,z^{D}\right) $, by the interval
\begin{equation}
ds^{2}=\left( \alpha /\tau \right) ^{2}[d\tau ^{2}-dr^{2}-r^{2}d\phi
^{2}-\left( d\mathbf{z}\right) ^{2}],  \label{ds2}
\end{equation}%
with the conformal time coordinate $\tau $, $-\infty <\tau <0$. The
corresponding synchronous time $t$ is expressed as $t=-\alpha \ln
(\left\vert \tau \right\vert /\alpha ),\ -\infty <t<+\infty $. For the
remaining coordinates we assume that $0\leqslant r<\infty $, $0\leqslant
\phi \leqslant \phi _{0}$, $-\infty <z^{l}<+\infty $, $l=3,\ldots ,D$. For $%
\phi _{0}=2\pi $ the geometry is reduced to the standard dS one given in
inflationary coordinates. In the case $\phi _{0}<2\pi $, though the local
geometrical characteristics for $r\neq 0$ remain the same, the global
properties are different. The special case $D=3$ corresponds to a straight
cosmic string with the core along the axis $z^{3}$ and with the planar angle
deficit $2\pi -$ $\phi _{0}$ determined by the linear mass density of the
string. In \cite{Ghez02} it has been shown that the vortex solution of the
Einstein-Abelian-Higgs equations in the presence of a cosmological constant
induces a deficit angle into dS spacetime. The cosmological constant $%
\Lambda $ is expressed in terms of the parameter $\alpha $ in the line
element (\ref{ds2}) by the relation $\Lambda =D(D-1)/(2\alpha ^{2})$.

The presence of the angle deficit gives rise to a number of interesting
topological effect in quantum field theory. Here we are interested in the
influence of the cosmic string on the vacuum fluctuations of the
electromagnetic field. The properties of these fluctuations are encoded in
the two-point functions which describe the correlations of the fluctuations
at different spacetime points. These correlators are VEVs of bilinear
combinations of the vector potential operator $A_{\mu }(x)$, where $x=(\tau
,r,\phi ,\mathbf{z})$ stands for the spacetime point. By expanding this
operator in terms of a complete set $\{A_{(\beta )\mu },A_{(\beta )\mu
}^{\ast }\}$ of solutions to the classical Maxwell equations and by using
the definition of the vacuum state $|0\rangle $, we can see that for a given
bilinear combination $f(A_{\mu }(x),A_{\nu }(x^{\prime }))$ the
corresponding VEV is presented in the form of the mode sum
\begin{equation}
\langle 0|f(A_{\mu }(x),A_{\nu }(x^{\prime }))|0\rangle =\sum_{\beta
}f(A_{(\beta )\mu }(x),A_{(\beta )\nu }^{\ast }(x^{\prime })).
\label{ModeSum}
\end{equation}%
Here, the set of quantum numbers $\beta $ specifies the electromagnetic mode
functions and in the right-hand side $\sum_{\beta }$ is understood as a
summation over discrete quantum numbers and an integration over continuous
ones. Hence, as the first stage, we need to find the complete set of
cylindrical electromagnetic modes on dS bulk in the presence of the cosmic
string.

It is convenient to fix the gauge degrees of the freedom by the Coulomb
gauge with $A_{0}=0$ and $\partial _{l}(\sqrt{|g|}A^{l})=0$ for $l=1,...,D$.
For the metric tensor%
\begin{equation}
g_{\mu \nu }=\left( \alpha /\tau \right) ^{2}\mathrm{diag}%
(1,-1,-r^{2},-1,\ldots ,-1),  \label{metric}
\end{equation}
the latter equation is reduced to $\partial _{l}(rA^{l})=0$ and coincides
with the corresponding equation in the Minkowski bulk. The procedure to find
the complete set of solutions to the Maxwell equations is similar to that we
have already described in \cite{Saha16} for the bulk in the absence of the
cosmic string. The only difference is in the periodicity condition along the
azimuthal direction $\phi $. The corresponding part in the mode functions is
given by $e^{iqm\phi }$ with $q=2\pi /\phi _{0}$ and $m=0,\pm 1,\pm 2,\ldots
$. This leads to the dependence of the mode functions on the radial
coordinate in terms of the Bessel function $J_{q|m|}(\gamma r)$ with $%
0\leqslant \gamma <\infty $. The time-dependence appears in the form of the
linear combination of the functions $\eta ^{D/2-1}H_{D/2-1}^{(1)}(\omega
\eta )$ and $\eta ^{D/2-1}H_{D/2-1}^{(2)}(\omega \eta )$, where $\eta =|\tau
|=\alpha e^{-t/\alpha }$ and $H_{\nu }^{(l)}(y)$, $l=1,2$, are the Hankel
functions. The relative coefficient in the linear combination depends on the
choice of the vacuum state under consideration. Here we assume that the
field is prepared in the state that is the analog of the Bunch-Davies vacuum
state for a scalar field \cite{Bunc78}. For this state the coefficient of
the function $H_{D/2-1}^{(2)}(\omega \eta )$ is zero.

In $(D+1)$-dimensional spacetime, the electromagnetic field has $D-1$
polarization states. In what follows we specify them by the quantum number $%
\sigma =1,\ldots ,D-1$. For the polarization $\sigma =1$ the cylindrical
electromagnetic modes corresponding to the Bunch-Davies vacuum are presented
as%
\begin{equation}
A_{(\beta )\mu }(x)=c_{\beta }\eta ^{D/2-1}H_{D/2-1}^{(1)}(\omega \eta
)\left( 0,\frac{iqm}{r},-r\partial _{r},0,\ldots ,0\right) J_{q|m|}(\gamma
r)e^{iqm\phi +i\mathbf{k}\cdot \mathbf{z}},  \label{A1}
\end{equation}%
and for the polarizations $\sigma =2,\ldots ,D-1$ we get
\begin{equation}
A_{(\beta )\mu }(x)=c_{\beta }\omega \eta ^{D/2-1}H_{D/2-1}^{(1)}(\omega
\eta )\left( 0,\epsilon _{\sigma l}+i\frac{\mathbf{k}\cdot \mathbf{\epsilon }%
_{\sigma }}{\omega ^{2}}\partial _{l}\right) J_{q|m|}(\gamma r)e^{iqm\phi +i%
\mathbf{k}\cdot \mathbf{z}},  \label{A2}
\end{equation}%
with $l=1,\ldots ,D$. Here, $\mathbf{k}=(k_{3},\ldots ,k_{D})$, $\omega =%
\sqrt{\gamma ^{2}+k^{2}}$ and $k^{2}=\sum_{l=3}^{D}k_{l}^{2}$. For the
scalar products one has $\mathbf{k}\cdot \mathbf{z}=\sum_{l=3}^{D}k_{l}z^{l}$
and $\mathbf{k}\cdot \mathbf{\epsilon }_{\sigma
}=\sum_{l=3}^{D}k_{l}\epsilon _{\sigma l}$. The spatial components in (\ref%
{A1}) and (\ref{A2}) are given in cylindrical coordinates $(r,\phi ,\mathbf{z%
})$. For the components of the polarization vector we have $\epsilon
_{\sigma 1}=\epsilon _{\sigma 2}=0$, $\sigma =2,\ldots ,D-1$, and the
relations%
\begin{eqnarray}
\sum_{l,n=3}^{D}\left( \omega ^{2}\delta _{nl}-k_{l}k_{n}\right) \epsilon
_{\sigma l}\epsilon _{\sigma ^{\prime }n} &=&\gamma ^{2}\delta _{\sigma
\sigma ^{\prime }},  \notag \\
\omega ^{2}\sum_{\sigma =2}^{D-1}\epsilon _{\sigma n}\epsilon _{\sigma
l}-k_{n}k_{l} &=&\gamma ^{2}\delta _{nl},  \label{Pol}
\end{eqnarray}%
for $l,n=3,...,D$. The mode functions are specified by the set of quantum
numbers $\beta =(\gamma ,m,\mathbf{k},\sigma )$ and in (\ref{ModeSum})%
\begin{equation}
\sum_{\beta }=\sum_{\sigma =1}^{D-1}\sum_{m=-\infty }^{\infty }\int d\mathbf{%
k}\int_{0}^{\infty }d\gamma .\,  \label{Sumbet}
\end{equation}%
We have a single mode of the TE type ($\sigma =1$) and $D-2$ modes of the
TM\ type ($\sigma =2,\ldots ,D-1$).

The mode functions for vector fields are orthonormalized by the condition
\begin{equation}
\int d^{D}x\sqrt{|g|}g^{00}[A_{(\beta ^{\prime })\nu }^{\ast }(x)\nabla
_{0}A_{(\beta )}^{\nu }(x)-(\nabla _{0}A_{(\beta ^{\prime })\nu }^{\ast
}(x))A_{(\beta )}^{\nu }(x)]=4i\pi \delta _{\beta \beta ^{\prime }},
\label{NormCond}
\end{equation}%
where $\nabla _{\mu }$ stands for the covariant derivative and $\delta
_{\beta \beta ^{\prime }}$ is understood as the Kronecker symbol for
discrete components of the collective index $\beta $ ($m$ and $\sigma $) and
the Dirac delta function for the continuous ones ($\gamma $ and $\mathbf{k}$%
). From (\ref{NormCond}) for the normalization coefficient $c_{\beta }$ we
get
\begin{equation}
|c_{\beta }|^{2}=\frac{q}{4(2\pi \alpha )^{D-3}\gamma },  \label{Normc}
\end{equation}%
for all the polarizations $\sigma =1,\ldots ,D-1$.

The Minkowskian limit of the problem under consideration corresponds to $%
\alpha \rightarrow \infty $ for a fixed value of the proper time $t$. In
this limit one has $\eta =\alpha e^{-t/\alpha }\approx \alpha -t$ and, up to
the phase (that can be absorbed into the normalization coefficient $c_{\beta
}$), the function $\eta ^{D/2-1}H_{D/2-1}^{(1)}(\omega \eta )$ is reduced to
$\sqrt{2/(\pi \omega )}\alpha ^{(D-3)/2}e^{-i\omega t}$. As a result, from (%
\ref{A1}) and (\ref{A2}) one gets the corresponding mode functions for a
string in background of $(D+1)$-dimensional Minkowski spacetime. The case $%
D=3$ has been considered previously in \cite{Beze07}. The electromagnetic
field is conformally invariant in $D=3$ and the modes (\ref{A1}) and (\ref%
{A2}) coincide with the Minkowskain modes having the time dependence $%
e^{-i\omega \eta }$.

\section{Two-point functions}

\label{sec:2point}

We consider a free field theory (the only interaction is with the background
gravitational field) and all the information about the vacuum state is
encoded in two-point functions. Given the complete set of normalized mode
functions for the vector potential, we can evaluate the two-point function $%
\langle 0|A_{l}(x)A_{m}(x^{\prime })|0\rangle \equiv \langle
A_{l}(x)A_{m}(x^{\prime })\rangle $ for the electromagnetic field by using
the mode-sum formula (\ref{ModeSum}):
\begin{equation}
\langle A_{l}(x)A_{m}(x^{\prime })\rangle =\sum_{\beta }\,A_{(\beta
)l}\left( x\right) A_{(\beta )m}^{\ast }\left( x^{\prime }\right) ,
\label{AA0}
\end{equation}%
with $\sum_{\beta }$ from (\ref{Sumbet}). Substituting the functions (\ref%
{A1}), (\ref{A2}) and using the relation (\ref{Pol}), the two-point function
is presented in the form%
\begin{eqnarray}
\langle A_{l}(x)A_{p}(x^{\prime })\rangle  &=&\frac{q\left( \eta \eta
^{\prime }\right) ^{D/2-1}}{\pi ^{2}\left( 2\pi \alpha \right) ^{D-3}}%
\sum_{m=-\infty }^{\infty }e^{imq\Delta \phi }\int d\mathbf{k}\,e^{i\mathbf{k%
}\cdot \Delta \mathbf{z}}  \notag \\
&&\times \int_{0}^{\infty }d\gamma \,\frac{\gamma }{\omega ^{2}}%
K_{D/2-1}(e^{-i\pi /2}\eta \omega )K_{D/2-1}(e^{i\pi /2}\eta ^{\prime
}\omega )f_{lp}\left( k,\gamma ,r,r^{\prime }\right) ,  \label{AA1}
\end{eqnarray}%
where $\Delta \phi =\phi -\phi ^{\prime }$, $\Delta \mathbf{z}=\mathbf{z}-%
\mathbf{z}^{\prime }$ and instead of the Hankel function we have introduced
the Macdonald function $K_{\nu }(x)$. In (\ref{AA1}), the functions of the
radial coordinates are defined by the expressions%
\begin{eqnarray}
f_{11}\left( k,\gamma ,r,r^{\prime }\right)  &=&k^{2}J_{q|m|}^{\prime
}(\gamma r)J_{q|m|}^{\prime }(\gamma r^{\prime })+\left( k^{2}+\gamma
^{2}\right) \frac{q^{2}m^{2}}{\gamma ^{2}rr^{\prime }}J_{q|m|}(\gamma
r)J_{q|m|}(\gamma r^{\prime }),  \notag \\
f_{12}\left( k,\gamma ,r,r^{\prime }\right)  &=&-i\frac{qm}{\gamma r}\left[
rk^{2}J_{q|m|}^{\prime }(\gamma r)J_{q|m|}(\gamma r^{\prime })+r^{\prime
}\omega ^{2}J_{q|m|}^{\prime }(\gamma r^{\prime })J_{q|m|}(\gamma r)\right] ,
\notag \\
f_{22}\left( k,\gamma ,r,r^{\prime }\right)  &=&\frac{q^{2}m^{2}}{\gamma ^{2}%
}k^{2}J_{q|m|}(\gamma r)J_{q|m|}(\gamma r^{\prime })+rr^{\prime }\omega
^{2}J_{q|m|}^{\prime }(\gamma r)J_{q|m|}^{\prime }(\gamma r^{\prime }),
\label{f22}
\end{eqnarray}%
and%
\begin{eqnarray}
f_{1l}\left( k,\gamma ,r,r^{\prime }\right)  &=&ik_{l}\gamma
J_{q|m|}^{\prime }(\gamma r)J_{q|m|}(\gamma r^{\prime }),  \notag \\
f_{2l}\left( k,\gamma ,r,r^{\prime }\right)  &=&-qmk_{l}J_{q|m|}(\gamma
r)J_{q|m|}(\gamma r^{\prime }),  \notag \\
f_{lp}\left( k,\gamma ,r,r^{\prime }\right)  &=&\left( \omega ^{2}\delta
_{lp}-k_{l}k_{p}\right) J_{q|m|}(\gamma r)J_{q|m|}(\gamma r^{\prime }),
\label{flp}
\end{eqnarray}%
with $l,p=3,\ldots ,D-1$. The remaining nonzero components are found by
using the relation%
\begin{equation}
f_{lp}\left( k,\gamma ,r,r^{\prime }\right) =f_{pl}^{\ast }\left( k,\gamma
,r^{\prime },r\right) .  \label{flprel}
\end{equation}

Having the two-point functions we can evaluate the VEVs of the squared
electric and magnetic fields. For the VEV\ of the squared electric field one
has
\begin{equation}
\langle E^{2}\rangle =\lim_{x^{\prime }\rightarrow x}C_{E}(x,x^{\prime }),
\label{E2}
\end{equation}%
where the corresponding two-point function is expressed as%
\begin{equation}
C_{E}(x,x^{\prime })=-g^{00^{\prime }}(x,x^{\prime })g^{lp^{\prime
}}(x,x^{\prime })\partial _{0}\partial _{0}^{\prime }\langle
A_{l}(x)A_{p}(x^{\prime })\rangle ,  \label{CE}
\end{equation}%
with the parallel propagator $g^{\mu \nu ^{\prime }}(x,x^{\prime })$. For
the geometry under consideration the nonzero components of the latter are
given by%
\begin{eqnarray}
g^{00^{\prime }}(x,x^{\prime }) &=&-g^{ll^{\prime }}(x,x^{\prime })=\frac{%
\eta \eta ^{\prime }}{\alpha ^{2}},  \notag \\
g^{11^{\prime }}(x,x^{\prime }) &=&rr^{\prime }g^{22^{\prime }}(x,x^{\prime
})=-\frac{\eta \eta ^{\prime }}{\alpha ^{2}}\cos \Delta \phi ,  \notag \\
rg^{21^{\prime }}(x,x^{\prime }) &=&-r^{\prime }g^{12^{\prime }}(x,x^{\prime
})=\frac{\eta \eta ^{\prime }}{\alpha ^{2}}\sin \Delta \phi ,  \label{gmunup}
\end{eqnarray}%
where $l=3,\ldots ,D$.

By taking into account the representation (\ref{AA1}) we find the expression%
\begin{eqnarray}
C_{E}(x,x^{\prime }) &=&\frac{8q\left( \eta \eta ^{\prime }\right) ^{D/2+1}}{%
\left( 2\pi \right) ^{D-1}\alpha ^{D+1}}\sideset{}{'}{\sum}_{m=0}^{\infty
}\left\{ \cos (mq\Delta \phi )\left[ (D-2)\mathcal{J}_{D/2-2}^{(0,2)}+(D-3)%
\mathcal{J}_{D/2-2}^{(1,1)}\right] \right.   \notag \\
&&+\left[ \cos (mq\Delta \phi )\cos \Delta \phi \left( \partial _{r}\partial
_{r^{\prime }}+\frac{q^{2}m^{2}}{rr^{\prime }}\right) +\frac{qm}{rr^{\prime }%
}\sin (mq\Delta \phi )\sin \Delta \phi \right.   \notag \\
&&\left. \times \left. \left( r\partial _{r}+r^{\prime }\partial _{r^{\prime
}}\right) \right] \left( \mathcal{J}_{D/2-2}^{(0,1)}+2\mathcal{J}%
_{D/2-2}^{(1,0)}\right) \right\} ,  \label{CE1}
\end{eqnarray}%
where
\begin{equation}
\mathcal{J}_{\nu }^{(n,p)}=\int d\mathbf{k}\,e^{i\mathbf{k}\cdot \Delta
\mathbf{z}}\int_{0}^{\infty }d\gamma \,k^{2n}\gamma ^{2p-1}K_{\nu }(e^{-i\pi
/2}\omega \eta )K_{\nu }(e^{i\pi /2}\omega \eta ^{\prime })J_{qm}(\gamma
r)J_{qm}(\gamma r^{\prime }).  \label{Jnp}
\end{equation}%
The prime on the summation sign in (\ref{CE1}) means that the term $m=0$
should be taken with an additional coefficient 1/2. The integrals (\ref{Jnp}%
) for $n=0,1$ and $p=0,1,2$ are evaluated in Appendix. By using the
corresponding results (\ref{Jnp3}), (\ref{J10a}) and (\ref{J10b}), the
correlator is presented as%
\begin{eqnarray}
C_{E}(x,x^{\prime }) &=&\frac{16q\left( \eta \eta ^{\prime }\right) ^{D/2+1}%
}{\pi ^{D/2}\alpha ^{D+1}}\int_{0}^{\infty }du\,u^{D/2}e^{u(\eta ^{2}+\eta
^{\prime 2}-|\Delta \mathbf{z}|^{2})}K_{D/2-2}(2\eta \eta ^{\prime }u)
\notag \\
&&\times \left\{ \left[ \partial _{w}w+2\left( D/2-1-|\Delta \mathbf{z}%
|^{2}u\right) \right] \left[ \cos \Delta \phi \left( \partial _{w}+b\right) -%
\frac{1}{w}\sin \Delta \phi \partial _{\Delta \phi }\right] \right.   \notag
\\
&&\left. +(D-2)\partial _{w}w+(D-3)\left( D/2-1-|\Delta \mathbf{z}%
|^{2}u\right) \right\} \sideset{}{'}{\sum}_{m=0}^{\infty }\cos (mq\Delta
\phi )e^{-bw}I_{qm}(w),  \label{CE2}
\end{eqnarray}%
with the notations
\begin{equation}
w=2rr^{\prime }u,\;b=\frac{r^{2}+r^{\prime 2}}{2rr^{\prime }}.  \label{wbn}
\end{equation}

For the further transformation of the expression (\ref{CE2}) we use the
formula \cite{Beze15,Beze12}%
\begin{equation}
\sideset{}{'}{\sum}_{m=0}^{\infty }\cos (qm\Delta \phi )I_{qm}\left(
w\right) =\frac{1}{2q}\sum_{l}e^{w\cos (2l\pi /q-\Delta \phi )}-\frac{1}{%
4\pi }\sum_{j=\pm 1}\int_{0}^{\infty }dy\frac{\sin (q\pi +jq\Delta \phi
)e^{-w\cosh y}}{\cosh (qy)-\cos (q\pi +jq\Delta \phi )},  \label{Summ}
\end{equation}%
where the summation in the first term on the right-hand side goes under the
condition
\begin{equation}
-q/2+q\Delta \phi /(2\pi )\leqslant l\leqslant q/2+q\Delta \phi /(2\pi ).
\label{krange}
\end{equation}%
If $-q/2+q\Delta \phi /(2\pi )$ or $q/2+q\Delta \phi /(2\pi )$ are integers,
then the corresponding terms in the first sum on the right-hand side of (\ref%
{Summ}) should be taken with the coefficient 1/2. The application of (\ref%
{Summ}) leads to the expression%
\begin{equation}
C_{E}(x,x^{\prime })=C_{E}^{(1)}(x,x^{\prime })+\sin \Delta \phi \partial
_{\Delta \phi }C_{E}^{(2)}(x,x^{\prime }).  \label{CEn}
\end{equation}%
Here and below we use the notation%
\begin{eqnarray}
C_{J}^{(i)}(x,x^{\prime }) &=&\frac{8\left( \eta \eta ^{\prime }\right)
^{D/2+1}}{\pi ^{D/2}\alpha ^{D+1}}\left[ \sum_{l}g_{J}^{(i)}(x,x^{\prime
},-\cos (2l\pi /q-\Delta \phi ))\right.  \notag \\
&&\left. -\frac{q}{2\pi }\sum_{j=\pm 1}\int_{0}^{\infty }dy\frac{\sin (q\pi
+jq\Delta \phi )g_{J}^{(i)}(x,x^{\prime },\cosh y)}{\cosh (qy)-\cos (q\pi
+jq\Delta \phi )}\right] ,  \label{CEi}
\end{eqnarray}%
for $i=1,2$ and $J=E,M$. The function with $J=M$ will appear in the
expression for the VEV of the squared magnetic field. The functions $%
g_{J}^{(i)}(x,x^{\prime },y)$ in (\ref{CEi}) have the representation
\begin{equation}
g_{J}^{(i)}(x,x^{\prime },y)=\int_{0}^{\infty }du\,u^{D/2}e^{u(\eta
^{2}+\eta ^{\prime 2}-|\Delta \mathbf{z}|^{2}-r^{2}-r^{\prime 2}-2rr^{\prime
}y)}K_{\nu _{J}}(2\eta \eta ^{\prime }u)h_{J}^{(i)}(y,u),  \label{gJ}
\end{equation}%
where
\begin{equation}
\nu _{J}=\left\{
\begin{array}{cc}
D/2-2, & J=E \\
D/2-1, & J=M%
\end{array}%
\right. .  \label{nuJ}
\end{equation}%
For the electric field, the functions in the integrand of (\ref{gJ}) are
given by the expressions
\begin{eqnarray}
h_{E}^{(1)}(y,u) &=&(D-2-y\cos \Delta \phi )\left[ 1-u\left( r^{2}+r^{\prime
2}+2rr^{\prime }y\right) \right]  \notag \\
&&+(D-3-2y\cos \Delta \phi )\left( D/2-1-|\Delta \mathbf{z}|^{2}u\right) ,
\notag \\
h_{E}^{(2)}(y,u) &=&\frac{1}{2rr^{\prime }u}\left[ u\left( r^{2}+r^{\prime
2}+2rr^{\prime }y\right) -2\left( D/2-1-|\Delta \mathbf{z}|^{2}u\right) %
\right] .  \label{hE2}
\end{eqnarray}%
The functions $h_{M}^{(i)}(y,u)$ for the magnetic field will be defined
below.

The contribution of the $l=0$ term in (\ref{CEi}) to the function (\ref{CEn}%
) corresponds to the correlator in dS spacetime in the absence of the cosmic
string (for the two-point functions of vector fields, including the massive
ones, see \cite{Alle86}). It is simplified to%
\begin{equation}
C_{E}^{\mathrm{(dS)}}(x,x^{\prime })=\frac{2(D-1)}{(2\pi )^{D/2}\alpha ^{D+1}%
}\int_{0}^{\infty }du\,u^{D/2}e^{uZ(x,x^{\prime })}\left( D-u\frac{|\Delta
\mathbf{x}|^{2}}{\eta \eta ^{\prime }}\right) K_{D/2-2}(u),  \label{CEq11}
\end{equation}%
where $|\Delta \mathbf{x}|^{2}=r^{2}+r^{\prime 2}-2rr^{\prime }\cos \Delta
\phi +|\Delta \mathbf{z}|^{2}$ is the square of the spatial distance between
the points $x$ and $x^{\prime }$ and we have defined the dS invariant
quantity%
\begin{equation}
Z(x,x^{\prime })=1+\frac{\left( \Delta \eta \right) ^{2}-|\Delta \mathbf{x}%
|^{2}}{2\eta \eta ^{\prime }}.  \label{Z}
\end{equation}%
For the latter one has $Z(x,x^{\prime })=\cos [\sigma (x,x^{\prime })/\alpha
]$, with $\sigma (x,x^{\prime })$ being the proper distance along the
shortest geodesic connecting the points $x$ and $x^{\prime }$ if they are
spacelike separated. The integral in (\ref{CEq11}) is expressed in terms of
the hypergeometric function. Separating the $l=0$ terms in the expressions
for $C_{E}^{(1)}(x,x^{\prime })$ and $C_{E}^{(2)}(x,x^{\prime })$, the
remaining part in (\ref{CEn}) corresponds to the contribution induced by the
presence of the cosmic string.

For points $x$ and $x^{\prime }$ close to each other, the dominant
contribution to the integral in (\ref{CEq11}) comes from large values of $u$
and we can use the corresponding asymptotic for the function $K_{D/2-2}(u)$.
To the leading order, for the pure dS part this gives%
\begin{equation}
C_{E}^{\mathrm{(dS)}}(x,x^{\prime })\approx \frac{2(D-1)\Gamma ((D+1)/2)}{%
\pi ^{(D-1)/2}\sigma ^{D+1}(x,x^{\prime })}\left[ D-\frac{(D+1)|\Delta
\mathbf{x}|^{2}}{|\Delta \mathbf{x}|^{2}-\left( \Delta \eta \right) ^{2}}%
\right] .  \label{CEq1cl}
\end{equation}%
In this limit the effects of the background curvature are small. Note that,
for points outside the cosmic string core, $r\neq 0$, the divergences in the
coincidence limit of $C_{E}(x,x^{\prime })$ are contained in the pure dS
part $C_{E}^{\mathrm{(dS)}}(x,x^{\prime })$ only. This is related to the
fact that in our simplified model the presence of the cosmic string does not
change the local geometry at those points.

\section{VEV of the squared electric field}

\label{sec:E2}

The VEV of the squared electric field is obtained from (\ref{CEn}) in the
coincidence limit. Separating the pure dS part $C_{E}^{\mathrm{(dS)}%
}(x,x^{\prime })$, the remaining topological contribution is finite in that
limit for $r\neq 0$. Consequently, the renormalization is reduced to the one
in dS spacetime. The contribution of the last term in (\ref{CEn}) to the
cosmic string induced part in the VEV of the field squared vanishes. As a
result, the VEV\ of the squared electric field is presented in the
decomposed form%
\begin{equation}
\langle E^{2}\rangle =\langle E^{2}\rangle _{\mathrm{dS}}+\frac{8\alpha
^{-D-1}}{(2\pi )^{D/2}}\left[ \sum_{l=1}^{[q/2]}g_{E}(r/\eta ,s_{l})-\frac{q%
}{\pi }\sin (q\pi )\int_{0}^{\infty }dy\frac{g_{E}(r/\eta ,\cosh y)}{\cosh
(2qy)-\cos (q\pi )}\right] ,  \label{E21}
\end{equation}%
where $[q/2]$ is the integer part of $q/2$. In (\ref{E21}), $\langle
E^{2}\rangle _{\mathrm{dS}}$ is the renormalized VEV in the absence of the
cosmic string and the remaining part is induced by the cosmic string
(topological part). Here and in what follows we use the notation $s_{l}=\sin
(\pi l/q)$ and
\begin{equation}
g_{E}(x,y)=\int_{0}^{\infty }du\,u^{D/2}K_{D/2-2}(u)e^{u-2x^{2}y^{2}u}\left[
2ux^{2}y^{2}\left( 2y^{2}-D+1\right) +\left( D-1\right) \left(
D/2-2y^{2}\right) \right] .  \label{gE1}
\end{equation}%
If the parameter $q$ is equal to an even integer the term $l=q/2$ in (\ref%
{E21}) should be taken with an additional coefficient 1/2. The VEV (\ref{E21}%
) depends on $r$ and $\eta $ in the form of the combination $r/\eta $. The
latter property is a consequence of the maximal symmetry of dS spacetime.
Note that, for a given $\eta $, the ratio $\alpha r/\eta $ is the proper
distance from the string. Hence, $r/\eta $ is the proper distance measured
in units of the dS curvature scale $\alpha $. From the maximal symmetry of
dS spacetime and of the Bunch-Davies vacuum state we expect that the pure dS
part does not depend on the spacetime point and $\langle E^{2}\rangle _{%
\mathrm{dS}}=\mathrm{const}\cdot \alpha ^{-D-1}$.

For odd values of $D$ the integral in (\ref{gE1}) is expressed in terms of
elementary functions. In particular, for $D=3$ and $D=5$ one has%
\begin{eqnarray}
g_{E}(x,y) &=&-\sqrt{\frac{\pi }{2}}\frac{1}{4x^{4}y^{4}},\;D=3,  \notag \\
g_{E}(x,y) &=&-\sqrt{\frac{\pi }{2}}\frac{1+y^{2}}{2x^{6}y^{6}},\;D=5.
\label{gE35}
\end{eqnarray}%
In these cases, the topological part in the squared electric field is
written in terms of the function%
\begin{equation}
c_{n}(q)=\sum_{l=1}^{[q/2]}s_{l}^{-n}-\frac{q}{\pi }\sin (q\pi
)\int_{0}^{\infty }dy\frac{\cosh ^{-n}y}{\cosh (2qy)-\cos (q\pi )}.
\label{cnq}
\end{equation}%
For even $n$, this function can be found by using the recurrence scheme
described in \cite{Beze06}. In particular, one has $c_{2}(q)=(q^{2}-1)/6$
and
\begin{eqnarray}
c_{4}(q) &=&\frac{q^{2}-1}{90}\left( q^{2}+11\right) ,  \notag \\
c_{6}(q) &=&\frac{q^{2}-1}{1890}(2q^{4}+23q^{2}+191).  \label{c6}
\end{eqnarray}%
As a result, the corresponding VEVs are presented as%
\begin{equation}
\langle E^{2}\rangle =\langle E^{2}\rangle _{\mathrm{dS}}-\frac{\left(
q^{2}-1\right) \left( q^{2}+11\right) }{180\pi (\alpha r/\eta )^{4}},
\label{E23}
\end{equation}%
for $D=3$ and%
\begin{equation}
\langle E^{2}\rangle =\langle E^{2}\rangle _{\mathrm{dS}}-\frac{\left(
q^{2}-1\right) \left( q^{4}+22q^{2}+211\right) }{1890\pi ^{2}\left( \alpha
r/\eta \right) ^{6}},  \label{E25}
\end{equation}%
for $D=5$. In the case $D=3$ the electromagnetic field is conformally
invariant and the topological part in (\ref{E23}) is obtained from the
corresponding result for a cosmic string in Minkowski bulk by the standard
conformal transformation. The latter is reduced to the multiplication of the
Minkowskian result by the factor $(\eta /\alpha )^{4}$.

As it has been mentioned before, the Minkowskian limit corresponds to $%
\alpha \rightarrow \infty $ for a fixed value of the time coordinate $t$. In
this case one has $\eta \approx \alpha -t$ and $\eta $ is large. Hence, we
need the asymptotic of the function (\ref{gE1}) for small values of $x$. In
this limit the dominant contribution to the integral comes from large values
of $u$ and using the asymptotic expression for the Macdonald function for
large argument, to the leading order we find%
\begin{equation}
g_{E}(x,y)\approx -\sqrt{\pi }\frac{\Gamma \left( \left( D+1\right)
/2\right) }{2^{D/2+2}x^{D+1}y^{D+1}}\left[ 2\left( D-3\right) y^{2}+D-1%
\right] .  \label{gEM}
\end{equation}%
As a consequence, for a string in the Minkowski bulk one gets%
\begin{equation}
\langle E^{2}\rangle ^{(M)}=-\frac{2\Gamma \left( \left( D+1\right)
/2\right) }{(4\pi )^{(D-1)/2}r^{D+1}}\left[ \left( D-3\right) c_{D-1}(q)+%
\frac{D-1}{2}c_{D+1}(q)\right] .  \label{E2M}
\end{equation}%
For $D=3$, this result is conformally related to the topological part in (%
\ref{E23}). It is of interest to note that, though the electromagnetic field
is not conformally invariant for $D=5$, the latter property is valid in this
case as well: $\langle E^{2}\rangle ^{(M)}=(\langle E^{2}\rangle -\langle
E^{2}\rangle _{\mathrm{dS}})\left( \alpha /\eta \right) ^{D+1}$, for $D=3,5$.

Now let us consider the asymptotic behavior of the VEV (\ref{E21}) at large
and small distances from the string. At large distances, $r/\eta \gg 1$, we
need the asymptotic expressions for the function $g_{E}(x,y)$ in the limit $%
x\gg 1$. In this limit the dominant contribution to the integral in (\ref%
{gE1}) comes from the region near the lower limit of the integration. By
using the asymptotic expression for the Macdonald function for small
argument, to the leading order we get%
\begin{equation}
g_{E}(x,y)\approx \frac{2^{D/2-5}}{y^{6}x^{6}}\Gamma \left( \frac{D}{2}%
-2\right) \left[ \left( D-1\right) \left( \frac{D}{2}-3\right) +2\left(
4-D\right) y^{2}\right] ,  \label{gEas1}
\end{equation}%
for $D>4$ and $g_{E}(x,y)\approx -3\ln (yx)/(2y^{6}x^{6})$ for $D=4$. In the
case $D>4$ this gives%
\begin{equation}
\langle E^{2}\rangle \approx \langle E^{2}\rangle _{\mathrm{dS}}+\frac{%
\Gamma \left( D/2-2\right) }{4\pi ^{D/2}\alpha ^{D+1}\left( r/\eta \right)
^{6}}\left[ 2\left( 4-D\right) c_{4}(q)+\left( D-1\right) \left( \frac{D}{2}%
-3\right) c_{6}(q)\right] ,  \label{E2far}
\end{equation}%
with the functions (\ref{c6}). Note that for $D=5$ the asymptotic (\ref%
{E2far}) coincides with the exact result (\ref{E25}). For $D=4$ the large
distance asymptotic is given by%
\begin{equation}
\langle E^{2}\rangle \approx \langle E^{2}\rangle _{\mathrm{dS}}-\frac{%
\left( q^{2}-1\right) \ln (r/\eta )}{630\pi ^{2}\alpha ^{5}(r/\eta )^{6}}%
(2q^{4}+23q^{2}+191).  \label{E2far2}
\end{equation}%
Hence, at large distances from the string, $r/\eta \gg 1$, the topological
part in the VEV\ of the electric field squared decays as $(\eta /r)^{4}$ for
$D=3$, as $\ln (r/\eta )(\eta /r)^{6}$ for $D=4$ and as $(\eta /r)^{6}$ for $%
D>4$. The pure dS part $\langle E^{2}\rangle _{\mathrm{dS}}$ is a constant
and it dominates in the total VEV\ at large distances. Note that at large
distances from the string the influence of the gravitational field on the
VEV is essential. In the Minkowskian case the decay of the VEV is as $%
1/r^{D+1}$ (see (\ref{E2M})) and depends on the number of spatial dimension.
For the dS bulk the VEV behaves as $1/r^{6}$ for all spatial dimensions $D>4$%
.

At proper distances from the string smaller than the dS curvature radius one
has $r/\eta \ll 1$ and the dominant contribution to the integral in (\ref%
{gE1}) comes from large values of $u$. The topological part dominates near
the string and by calculations similar to those for the Minkowskian limit we
get%
\begin{equation}
\langle E^{2}\rangle \approx (\eta /\alpha )^{D+1}\langle E^{2}\rangle
^{(M)},\;r/\eta \ll 1,  \label{E2near}
\end{equation}%
with $\langle E^{2}\rangle ^{(M)}$ given by (\ref{E2M}). This result is
natural because near the string the dominant contribution to the VEV comes
from the fluctuations with wavelengths smaller than the curvature radius and
the influence of the background gravitational field on the corresponding
modes is weak.

The VEV of the electric field squared determines the Casimir-Polder
interaction energy between the cosmic string and a neutral polarizable
microparticle placed close to the string, $U(r)=-\alpha _{P}\langle
E^{2}\rangle $, where $\alpha _{P}$ is the polarizability of the particle
(in the absence of dispersion). The correlators of the electromagnetic field
and the Casimir-Polder potential in the geometry of cosmic string on
background of $D=3$ Minkowski spacetime were investigated in \cite{Bard10}.

\section{Magnetic field correlators and VEV of the energy density}

\label{sec:LM}

As a next characteristic of the vacuum state we consider the VEV of the
Lagrangian density:%
\begin{equation}
\langle L\rangle =-\frac{1}{16\pi }g^{\mu \rho }g^{\nu \sigma }\langle
F_{\mu \nu }F_{\rho \sigma }\rangle .  \label{L}
\end{equation}%
Note that the quantity $g^{\mu \rho }g^{\nu \sigma }\langle F_{\mu \nu
}F_{\rho \sigma }\rangle $ is the Abelian analog of the gluon condensate in
quantum chromodynamics. The VEV (\ref{L}) is presented as the coincidence
limit
\begin{equation}
\langle L\rangle =\lim_{x^{\prime }\rightarrow x}C_{L}(x,x^{\prime }),
\label{L2}
\end{equation}%
with the corresponding correlator%
\begin{equation}
C_{L}(x,x^{\prime })=-\frac{1}{16\pi }g^{\mu \rho ^{\prime }}(x,x^{\prime
})g^{\nu \sigma ^{\prime }}(x,x^{\prime })\langle F_{\mu \nu }(x)F_{\rho
\sigma }(x^{\prime })\rangle .  \label{CL0}
\end{equation}%
The latter is decomposed into the electric and magnetic parts as
\begin{equation}
C_{L}(x,x^{\prime })=\frac{1}{8\pi }\left[ C_{E}(x,x^{\prime
})-C_{M}(x,x^{\prime })\right] ,  \label{CL}
\end{equation}%
where the magnetic part is given by the expression%
\begin{eqnarray}
C_{M}(x,x^{\prime }) &=&\frac{1}{2}g^{lm^{\prime }}(x,x^{\prime
})g^{np^{\prime }}(x,x^{\prime })\langle F_{ln}(x)F_{mp}(x^{\prime })\rangle
\notag \\
&=&\left[ g^{lm^{\prime }}(x,x^{\prime })g^{np^{\prime }}(x,x^{\prime
})-g^{nm^{\prime }}(x,x^{\prime })g^{lp^{\prime }}(x,x^{\prime })\right]
\partial _{l}\partial _{m^{\prime }}\langle A_{n}(x)A_{p^{\prime
}}(x^{\prime })\rangle .  \label{CM0}
\end{eqnarray}%
with the summation over the spatial indices $l,m,n,p=1,2,\ldots ,D$ (for a
scheme to measure the correlation functions for cosmological magnetic fields
based on TeV blazar observations see \cite{Tash13}).

By using (\ref{AA1}), after long calculations, the magnetic part is
presented in the form%
\begin{eqnarray}
C_{M}(x,x^{\prime }) &=&\frac{2q\left( \eta \eta ^{\prime }\right) ^{D/2+1}}{%
\pi ^{2}\left( 2\pi \right) ^{D-3}\alpha ^{D+1}}\sideset{}{'}{\sum}%
_{m=0}^{\infty }  \notag \\
&&\times \left\{ \left[ \cos (mq\Delta \phi )\cos \Delta \phi \left(
\partial _{r}\partial _{r^{\prime }}+\frac{q^{2}m^{2}}{rr^{\prime }}\right) +%
\frac{qm}{rr^{\prime }}\sin (mq\Delta \phi )\sin \Delta \phi \left(
r\partial _{r}+r^{\prime }\partial _{r^{\prime }}\right) \right] \right.
\notag \\
&&\left. \times \left( \left( D-2\right) \mathcal{J}_{D/2-1}^{(0,1)}+2%
\mathcal{J}_{D/2-1}^{(1,0)}\right) +\cos (mq\Delta \phi )\left[ \mathcal{J}%
_{D/2-1}^{(0,2)}+(D-3)\mathcal{J}_{D/2-1}^{(1,1)}\right] \right\} .
\label{CM1}
\end{eqnarray}%
By taking into account the representations for the functions $\mathcal{J}%
_{\nu }^{(n,p)}$ given in Appendix, for the correlator one gets%
\begin{eqnarray}
C_{M}(x,x^{\prime }) &=&\frac{16q\left( \eta \eta ^{\prime }\right) ^{D/2+1}%
}{\pi ^{D/2}\alpha ^{D+1}}\int_{0}^{\infty }du\,u^{D/2}e^{u(\eta ^{2}+\eta
^{\prime 2}-|\Delta \mathbf{z}|^{2})}K_{D/2-1}(2\eta \eta ^{\prime }u)
\notag \\
&&\times \left\{ \left[ (D-2)\partial _{w}w+2\left( D/2-1-|\Delta \mathbf{z}%
|^{2}u\right) \right] \left[ \cos \Delta \phi \left( \partial _{w}+b\right) -%
\frac{1}{w}\sin \Delta \phi \partial _{\Delta \phi }\right] \right.  \notag
\\
&&\left. +\partial _{w}w+(D-3)\left( D/2-1-|\Delta \mathbf{z}|^{2}u\right)
\right\} \sideset{}{'}{\sum}_{m=0}^{\infty }\cos (mq\Delta \phi
)e^{-bw}I_{qm}(w).  \label{CM2}
\end{eqnarray}

The further transformation is similar to that employed for the electric
field correlator. By using the formula (\ref{Summ}) we find%
\begin{equation}
C_{M}(x,x^{\prime })=C_{M}^{(1)}(x,x^{\prime })+\sin \Delta \phi \partial
_{\Delta \phi }C_{M}^{(2)}(x,x^{\prime }),  \label{CM3}
\end{equation}%
where the functions $C_{M}^{(i)}(x,x^{\prime })$ are defined in (\ref{CEi})
with $J=M$. In the corresponding definition the function $%
g_{M}^{(i)}(x,x^{\prime },y)$ is given by the expression (\ref{gJ}) with the
functions in the integrand%
\begin{eqnarray}
h_{M}^{(1)}(y,u) &=&\left[ 1-\left( D-2\right) y\cos \Delta \phi \right] %
\left[ 1-u\left( r^{2}+r^{\prime 2}+2rr^{\prime }y\right) \right]  \notag \\
&&+(D-3-2y\cos \Delta \phi )\left( D/2-1-|\Delta \mathbf{z}|^{2}u\right) ,
\label{hM1}
\end{eqnarray}%
and $h_{M}^{(2)}(y,u)=h_{E}^{(2)}(y,u)$. The contributions of the $l=0$
terms in (\ref{CEi}) to (\ref{CM3}) correspond to the correlator in dS
spacetime in the absence of the cosmic string ($q=1$):%
\begin{equation}
C_{M}^{\mathrm{(dS)}}(x,x^{\prime })=\frac{2(D-1)}{(2\pi )^{D/2}\alpha ^{D+1}%
}\int_{0}^{\infty }du\,u^{D/2}e^{uZ(x,x^{\prime })}\left( D-u\frac{|\Delta
\mathbf{x}|^{2}}{\eta \eta ^{\prime }}\right) K_{D/2-1}(u),  \label{CMq11}
\end{equation}%
For $D=3$ the electric and magnetic correlators coincide and, hence, the
correlator for the Lagrangian density vanishes. For close points $x$ and $%
x^{\prime }$, the leading term in the corresponding asymptotic expansion
coincides with that for the correlator of the electric field, $C_{M}^{%
\mathrm{(dS)}}(x,x^{\prime })\approx C_{E}^{\mathrm{(dS)}}(x,x^{\prime })$,
and is given by (\ref{CEq1cl}).

In (\ref{CM3}), separating the $l=0$ terms in the expressions (\ref{CEi})
for the functions $C_{M}^{(i)}(x,x^{\prime })$, the remaining part is the
contribution induced by the cosmic string. For $r\neq 0$, the latter is
finite in the coincidence limit. The renormalization is required for the
pure dS part only. Hence, for the VEV\ of the squared magnetic field,%
\begin{equation}
\langle B^{2}\rangle =\lim_{x^{\prime }\rightarrow x}C_{M}(x,x^{\prime }),
\label{B2lim}
\end{equation}%
one finds the decomposition%
\begin{equation}
\langle B^{2}\rangle =\langle B^{2}\rangle _{\mathrm{dS}}+\frac{8\alpha
^{-D-1}}{(2\pi )^{D/2}}\left[ \sum_{l=1}^{[q/2]}g_{M}(r/\eta ,s_{l})-\frac{q%
}{\pi }\sin (q\pi )\int_{0}^{\infty }dy\frac{g_{M}(r/\eta ,\cosh y)}{\cosh
(2qy)-\cos (q\pi )}\right] ,  \label{B2}
\end{equation}%
with the function%
\begin{eqnarray}
g_{M}(x,y) &=&\int_{0}^{\infty
}du\,u^{D/2}K_{D/2-1}(u)e^{u-2x^{2}y^{2}u}\left\{ (D-1)D/2\right.  \notag \\
&&\left. -4(D-2)y^{2}+2x^{2}y^{2}u\left[ 2(D-2)y^{2}-D+1\right] \right\} .
\label{gMn}
\end{eqnarray}%
Similar to (\ref{E21}), if $q/2$ is an integer, the term $l=q/2$ in (\ref{B2}%
) should be taken with an additional coefficient 1/2. Note that for $D>3$
the magnetic part of the field tensor is not a spatial vector. In (\ref{B2}%
), $\langle B^{2}\rangle _{\mathrm{dS}}$ is the corresponding renormalized
quantity in the absence of the cosmic string and, because of the maximal
symmetry of dS spacetime, does not depend on the spacetime point. From the
dimensional arguments we expect that $\langle B^{2}\rangle _{\mathrm{dS}}=%
\mathrm{const}\cdot \alpha ^{-D-1}$. For $D=3$, the VEV of the squared
magnetic field has been investigated in \cite{Camp13} by using the adiabatic
renormalization procedure. In this special case $\langle B^{2}\rangle _{%
\mathrm{dS}}=19/(40\pi \alpha ^{4})$ (note the different units used here and
in \cite{Camp13}).

For odd $D$, the function $g_{M}(x,y)$ is expressed in terms of the
elementary functions. In particular, for $D=3$ it coincides with $g_{E}(x,y)$%
, given by (\ref{gE35}), and for $D=5$ one has%
\begin{equation}
g_{M}(x,y)=\sqrt{\frac{\pi }{2}}\frac{\left( 3+x^{2}\right) y^{2}-1}{%
2x^{6}y^{6}},\;D=5.  \label{gM5}
\end{equation}%
In the latter case, the VEV of the squared magnetic field is presented as%
\begin{equation}
\langle B^{2}\rangle =\langle B^{2}\rangle _{\mathrm{dS}}+\frac{\left(
3+r^{2}/\eta ^{2}\right) c_{4}(q)-c_{6}(q)}{2\pi ^{2}\left( \alpha r/\eta
\right) ^{6}},  \label{B25}
\end{equation}%
where the functions $c_{4}(q)$ and $c_{6}(q)$ are defined in (\ref{c6}).

Let us consider the asymptotic behavior of the VEV (\ref{B2}) at large and
small distances from the string. If the proper distance from the string is
much smaller than the curvature radius of the dS spacetime one has $r/\eta
\ll 1$. For small $x$ the dominant contribution to (\ref{gMn}) comes from
large values of $u$. By using the corresponding asymptotic for the Macdonald
function, to the leading order we get%
\begin{equation}
\langle B^{2}\rangle \approx \left( \eta /\alpha \right) ^{D+1}\langle
B^{2}\rangle ^{(M)},  \label{B2near}
\end{equation}%
where%
\begin{equation}
\langle B^{2}\rangle ^{(M)}=\frac{2\Gamma \left( \left( D+1\right) /2\right)
}{(4\pi )^{(D-1)/2}r^{D+1}}\left[ \left( D-3\right) (D-2)c_{D-1}(q)-\frac{D-1%
}{2}c_{D+1}(q)\right] ,  \label{B2M}
\end{equation}%
is the corresponding VEV for the cosmic string in Minkowski bulk. In
particular, for $D=3$ one has $\langle B^{2}\rangle ^{(M)}=\langle
E^{2}\rangle ^{(M)}$ with $\langle E^{2}\rangle ^{(M)}$ given by the last
term in the right-hand side of (\ref{E23}).

At large distances from the string, $r/\eta \gg 1$, we need the asymptotic
of the function $g_{M}(x,y)$ for large $x$. In this limit, the dominant
contribution to the integral in (\ref{gMn}) comes from the region near the
lower limit of the integration and for the leading term one finds%
\begin{equation}
g_{M}(x,y)\approx \frac{2^{D/2-5}}{y^{4}x^{4}}(D-1)\left( D-4\right) \Gamma
(D/2-1).  \label{gMlarge}
\end{equation}%
For $D=4$ the leading term vanishes and we need to consider the next to the
leading contribution:%
\begin{equation}
g_{M}(x,y)\approx \frac{y^{2}-3/4}{y^{6}x^{6}}.  \label{gMD4}
\end{equation}%
By taking into account (\ref{c6}) and (\ref{gMlarge}), at distances $r/\eta
\gg 1$ one gets%
\begin{equation}
\langle B^{2}\rangle \approx \langle B^{2}\rangle _{\mathrm{dS}}+\frac{%
(D-1)\left( D-4\right) \Gamma (D/2-1)}{360\pi ^{D/2}\alpha ^{D+1}(r/\eta
)^{4}}\left( q^{2}-1\right) \left( q^{2}+11\right) ,  \label{B2large}
\end{equation}%
for $D\neq 4$ and
\begin{equation}
\langle B^{2}\rangle \approx \langle B^{2}\rangle _{\mathrm{dS}}+\frac{%
4c_{4}(q)-3c_{6}(q)}{2\pi ^{2}\alpha ^{5}\left( r/\eta \right) ^{6}},
\label{B2largeD4}
\end{equation}%
for $D=4$. In the special case $D=3$, the asymptotic (\ref{B2large})
coincides with the exact result.

\begin{figure}[tbph]
\begin{center}
\epsfig{figure=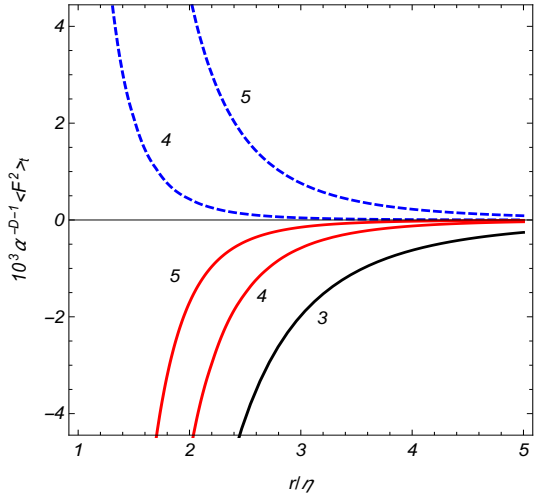,width=7.cm,height=6.5cm}
\end{center}
\caption{Topological contributions in the VEVs of the squared electric and
magnetic fields for $q=2.5$ and for spatial dimensions $D=3,4,5$ (the
numbers near the curves). The full/dashed curves correspond to the
electric/magnetic fields.}
\label{fig1}
\end{figure}

In figure \ref{fig1} we have plotted the topological contributions in the
VEVs of the squared electric and magnetic fields, $\langle F^{2}\rangle
_{t}=\langle F^{2}\rangle -\langle F^{2}\rangle _{\mathrm{dS}}$, $F=E,B$,
for $q=2.5$ and for spatial dimensions $D=3,4,5$ (the numbers near the
curves). The full/dashed curves correspond to the electric/magnetic fields.
In the case $D=3$ one has $\langle E^{2}\rangle _{t}=\langle B^{2}\rangle
_{t}$. Note that for $D=4,5$ the VEVs of the squared electric and magnetic
fields have opposite signs. The Casimir-Polder forces acting on a
polarizable particle are attractive.

Among the most interesting features of the inflation is the transition from
quantum to classical behavior of quantum fluctuations during the
quasiexponential expansion of the universe. An important example of this
type of effects is the classicalization of the vacuum fluctuations of the
inflaton field which underlies the most popular models of generation of
large-scale structure in the universe. A similar effect of classicalization
should take place for the electromagnetic fluctuations. In \cite{Camp13},
the quantum-to-classical transition of super-Hubble magnetic modes during
inflation has been considered as a possible mechanism for the generation of
galactic and galaxy cluster magnetic fields (see also \cite{Durr13} for the
further discussion). As it has been discussed above, the presence of cosmic
string induces shifts in the VEVs of the squared electric and magnetic
fields. As a consequence of the quantum-to-classical transition of the
corresponding fluctuations during the dS expansion, after inflation these
shifts will be imprinted as classical stochastic fluctuations of the
electric and magnetic fields surrounding the cosmic string. In the post
inflationary radiation dominated era the conductivity is high and the
currents in the cosmic plasma eliminate the electric fields whereas the
magnetic counterparts are frozen. As a consequence, the cosmic strings will
be surrounded by large scale magnetic fields. These fields would be among
the distinctive features of the cosmic strings produced during the inflation
and also of the corresponding inflationary models. Note that various types
of mechanisms for the generation of primordial magnetic fields from cosmic
strings in the post-inflationary era have been discussed in the literature
(see, for instance, \cite{Vach91}). For cosmic strings carrying a nonzero
magnetic flux in the core, azimuthal currents for charged fields are
generated around the string (see \cite{Beze15} and references therein).
These currents provide another mechanism for the generation of magnetic
fields by the cosmic strings.

Having the VEVs for the squared electric and magnetic fields, we can find
the VEV of the energy density $\varepsilon $ as%
\begin{equation}
\langle \varepsilon \rangle =\frac{\langle E^{2}\rangle +\langle
B^{2}\rangle }{8\pi }.  \label{eps}
\end{equation}%
It is decomposed into the pure dS part, $\langle \varepsilon \rangle _{%
\mathrm{dS}}$, and the topological contribution:%
\begin{equation}
\langle \varepsilon \rangle =\langle \varepsilon \rangle _{\mathrm{dS}}+%
\frac{2\alpha ^{-D-1}}{(2\pi )^{D/2+1}}\left[ \sum_{l=1}^{[q/2]}g_{0}(r/\eta
,s_{l})-\frac{q}{\pi }\sin (q\pi )\int_{0}^{\infty }dy\frac{g_{0}(r/\eta
,\cosh y)}{\cosh (2qy)-\cos (q\pi )}\right] ,  \label{eps1}
\end{equation}%
with $g_{0}(x,y)=g_{E}(x,y)+g_{M}(x,y)$. If $q$ is equal to an even integer,
the term $l=q/2$ in (\ref{eps1}) is taken with an additional coefficient
1/2. From the maximal symmetry of the dS spacetime it follows that $\langle
\varepsilon \rangle _{\mathrm{dS}}=\mathrm{const}/\alpha ^{D+1}$. In the
special case $D=3$ the latter is completely determined by the conformal
anomaly (see, for instance, \cite{Birr82}): $\langle \varepsilon \rangle _{%
\mathrm{dS}}=31/(480\pi ^{2}\alpha ^{4})$. Combining this with the result
for $\langle B^{2}\rangle _{\mathrm{dS}}$ we can find the VEV\ of the
squared electric field: $\langle E^{2}\rangle _{\mathrm{dS}}=1/(24\pi \alpha
^{4})$. In this case, the contributions of the electric and magnetic parts
to the topological term in the VEV of the energy density are the same and
the VEV is conformally related to the corresponding result on the Minkowski
bulk found in \cite{Frol87}.

Near the string, $r/\eta \ll 1$, the VEV of the energy density is dominated
by the topological part and to the leading order one has%
\begin{equation}
\langle \varepsilon \rangle \approx \frac{\Gamma \left( \left( D+1\right)
/2\right) }{(4\pi )^{(D+1)/2}(\alpha r/\eta )^{D+1}}\left[ \left( D-3\right)
^{2}c_{D-1}(q)-\left( D-1\right) c_{D+1}(q)\right] .  \label{epsNear}
\end{equation}%
The corresponding VEV\ in the Minkowski bulk is obtained from the right hand
side multiplying by the factor $(\alpha /\eta )^{D+1}$. Depending on the
parameters $D$ and $q$, the energy density (\ref{epsNear}) can be either
negative or positive. For $D=3$ it is always negative. At large distances
from the cosmic string and for $D>4$ the topological contribution in the
energy density is dominated by the magnetic part and decays as $(\eta
/r)^{4} $. For $D=4$ and at large distances the electric part dominates and
the energy density decays like $(\eta /r)^{6}\ln (r/\eta )$. In figure \ref%
{fig2} we display the dependence of the topological contribution in the
vacuum energy density as a function of the proper distance from the string
(measured in units of dS curvature scale). The graphs are plotted for
spatial dimensions $D=3,4,5$. As is seen, in general, the energy density is
not a monotonic function of the distance from the string. Moreover, in the
case $D=5$ the energy density changes the sign: it is negative near the
string (the electric part dominates) and positive at large distances from
the string (the magnetic contribution dominates).
\begin{figure}[tbph]
\begin{center}
\epsfig{figure=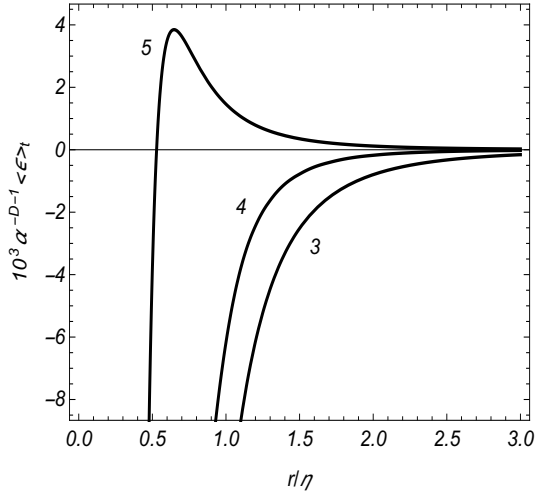,width=7.cm,height=6.5cm}
\end{center}
\caption{Topological contributions in the VEV of the energy density in
spatial dimensions $D=3,4,5$ (the numbers near the curves) for $q=2.5$.}
\label{fig2}
\end{figure}

\section{Conclusion}

\label{sec:Conc}

In the present paper we have investigated the influence of the cosmic string
on the vacuum fluctuations of the electromagnetic field in background of $%
(D+1)$-dimensional dS spacetime, assuming that the field is prepared in the
state which is the analog of the Bunch-Davies vacuum state for a scalar
field. In the problem under consideration the only interaction of the
quantum electromagnetic field is with the background gravitational field and
the information on the vacuum fluctuations is encoded in the two-point
functions. As such we have considered the Wightman function. For the
evaluation of the latter we have used the direct summation over the complete
set of electromagnetic cylindrical modes. The corresponding mode functions
for separate polarizations are given by (\ref{A1}) and (\ref{A2}).

Among the most important characteristics of the electromagnetic vacuum are
the VEVs of the squared electric and magnetic fields. The corresponding
two-point functions are given by (\ref{CEn}) and (\ref{CM2}), respectively,
with the function $C_{J}^{(i)}(x,x^{\prime })$ defined in (\ref{CEi}). One
of the advantages for these representations is that the contribution
corresponding to dS spacetime in the absence of the cosmic string is
explicitly extracted. In the model of the cosmic string under consideration
the local geometrical characteristics outside the string core are not
changed by the presence of the string. Consequently, the divergences and the
renormalization procedure for the VEVs are the same as those in pure dS
spacetime. The topological parts do not require a renormalization. The
renormalized VEVs for the squared electric and magnetic fields are presented
in the decomposed form, Eqs. (\ref{E21}) and (\ref{B2}), respectively, where
the first terms in the right-hand sides correspond to the renormalized VEVs
in dS spacetime in the absence of the cosmic string. As a consequence of the
maximal symmetry of dS spacetime and of the Bunch-Davies vacuum state these
contributions do not depend on the spacetime coordinates.

The topological parts in the VEVs depend on the time and the radial
coordinate through the ratio $r/\eta $ which presents the proper distance
from the string measured in units of the dS curvature radius. Near the
string, the dominant contribution to the VEVs comes from the fluctuations
with short wavelengths and the VEVs coincide with those for the string in
Minkowski bulk with the distance from the string replaced by the proper
distance $\alpha r/\eta $. The influence of the gravitational field on the
topological contributions in the VEVs is crucial at proper distances larger
than the curvature radius of the background geometry. This contribution in
the electric field squared decays as $(\eta /r)^{4}$ for $D=3$, as $\ln
(r/\eta )(\eta /r)^{6}$ for $D=4$ and as $(\eta /r)^{6}$ for $D>4$. For the
squared magnetic field the topological contribution decays as $(\eta /r)^{4}$
for $D\geqslant 3$. The exception is the case $D=4$ where the corresponding
coefficient vanishes and the next term in the expansion should be kept. In
this case the topological term falls off as $(\eta /r)^{6}$. In the
Minkowskian bulk the decay of the VEVs is as $1/r^{D+1}$ for both the
electric and magnetic fields.

The modifications of the electromagnetic field vacuum fluctuations during
the dS expansion phase, we have discussed here, will be imprinted in
large-scale stochastic perturbations of the electromagnetic fields
surrounding the cosmic string in the post-inflationary radiation dominated
era. The magnetic fields will be frozen in the cosmic plasma whereas the
electric fields will be eliminated by the induced currents.

We have also investigated the VEV of the electromagnetic energy density,
induced by a cosmic string. Near the string the topological contribution
dominates in the total VEV and the energy density behaves as $(\eta
/r)^{D+1} $. At distances from the string larger than the curvature radius
of the dS spacetime and for spatial dimensions $D>4$, the topological part
in the energy density is dominated by the magnetic contribution and decays
as $(\eta /r)^{4}$. For $D=4$ the electric field contribution dominates and
at large distances the string-induced energy density behaves as $(\eta
/r)^{6}\ln (r/\eta )$. For $D=3$ the topological contribution in the energy
density is negative and decays as $(\eta /r)^{4}$ for all distances. For
other spatial dimensions the energy density, in general, is not a monotonic
function of the distance from the string. For example, in the case $D=5$ the
energy density is negative near the string and positive at large distances.
It has a maximum for some intermediate value of the distance from the string.

\section*{Acknowledgments}

A. A. S. and N. A. S. were supported by the State Committee of Science
Ministry of Education and Science RA, within the frame of Grant No. SCS
15T-1C110, and by the Armenian National Science and Education Fund (ANSEF)
Grant No. hepth-4172.

\appendix

\section{Evaluation of the integrals}

\label{sec:App}

Here we describe the evaluation of the integrals (\ref{Jnp}) appearing in
the expressions of the two-point functions for the electromagnetic field. By
using the integral representation \cite{Wats66}%
\begin{equation}
K_{\nu }(e^{-i\pi /2}\eta \omega )K_{\nu }(e^{i\pi /2}\eta ^{\prime }\omega
)=\frac{1}{2}\int_{-\infty }^{+\infty }dy\,e^{-2\nu y}\int_{0}^{\infty }%
\frac{du}{u}e^{-u/2-\omega ^{2}\beta /(2u)},  \label{IntRepK}
\end{equation}%
with the notation
\begin{equation}
\beta =2\eta \eta ^{\prime }\cosh (2y)-\eta ^{2}-\eta ^{\prime 2},
\label{bet}
\end{equation}%
and redefining the integration variable $u/(2\beta )\rightarrow u$, one gets%
\begin{eqnarray}
\mathcal{J}_{\nu }^{(n,p)} &=&\frac{1}{2}\int d\mathbf{k}\,e^{i\mathbf{k}%
\cdot \Delta \mathbf{z}}\int_{-\infty }^{+\infty }dy\,e^{-2\nu
y}\int_{0}^{\infty }\frac{du}{u}e^{-u\beta }  \notag \\
&&\times \int_{0}^{\infty }d\gamma k^{2n}\gamma ^{2p-1}e^{-\omega
^{2}/4u}J_{qm}(\gamma r)J_{qm}(\gamma r^{\prime }).  \label{Jnp1}
\end{eqnarray}%
The integration over $\mathbf{k}$ is done with the help of the formula%
\begin{equation}
\int d\mathbf{k}\,k^{2n}e^{i\mathbf{k}\cdot \Delta \mathbf{z}%
-k^{2}/4u}=(4\pi )^{D/2-1}u^{D/2-1}\left[ 4u\left( D/2-1-|\Delta \mathbf{z}%
|^{2}u\right) \right] ^{n}e^{-|\Delta \mathbf{z}|^{2}u},  \label{Intk}
\end{equation}%
where $n=0,1$. Next, the integral over $y$ is expressed in terms of the
function $K_{\nu }(2\eta \eta ^{\prime }u)$ and we find%
\begin{eqnarray}
\mathcal{J}_{\nu }^{(n,p)} &=&\frac{1}{2}(4\pi )^{D/2-1}\int_{0}^{\infty
}du\,u^{D/2-2}\left[ 4u\left( D/2-1-|\Delta \mathbf{z}|^{2}u\right) \right]
^{n}e^{u\left( \eta ^{2}+\eta ^{\prime 2}-|\Delta \mathbf{z}|^{2}\right) }
\notag \\
&&\times K_{\nu }(2\eta \eta ^{\prime }u)\int_{0}^{\infty }d\gamma \gamma
^{2p-1}e^{-\gamma ^{2}/4u}J_{qm}(\gamma r)J_{qm}(\gamma r^{\prime }).
\label{Jnp2}
\end{eqnarray}%
For $p=1,2$ the integration over $\gamma $ is done by using the formula \cite%
{Prud86}%
\begin{equation}
\int_{0}^{\infty }d\gamma \gamma ^{2p-1}e^{-\gamma ^{2}/4u}J_{qm}(\gamma
r)J_{qm}(\gamma r^{\prime })=2\left( 4u^{2}\partial _{u}\right) ^{p-1}\left[
ug(r,r^{\prime },u)\right] ,  \label{gamInt}
\end{equation}%
with the function%
\begin{equation}
g(r,r^{\prime },u)=e^{-(r^{2}+r^{\prime 2})u}I_{qm}(2rr^{\prime }u).
\label{ge}
\end{equation}

For the evaluation of the integral in (\ref{Jnp2}) with $p=0$ we use the
integral representation
\begin{equation}
e^{-\gamma ^{2}/4u}=\gamma ^{2}\int_{1/4u}^{\infty }dt\,e^{-t\gamma ^{2}},
\label{IntRep}
\end{equation}
and then apply the formula (\ref{gamInt}) with $p=1$ for the $\gamma $%
-integral. In this way, we can see that
\begin{equation}
\int_{0}^{\infty }d\gamma \ \frac{e^{-\gamma ^{2}/4u}}{\gamma }J_{qm}(\gamma
r)J_{qm}(\gamma r^{\prime })=\frac{1}{2}\int_{0}^{u}\frac{dx}{x}%
\,g(r,r^{\prime },x).  \label{Intgam2}
\end{equation}%
The corresponding integral in the two-point function (\ref{CE1}) is acted by
the operators with the results%
\begin{equation}
\left( r\partial _{r}+r^{\prime }\partial _{r^{\prime }}\right) \int_{0}^{u}%
\frac{dx}{x}\,g(r,r^{\prime },x)=2e^{-bw}I_{qm}(w),  \label{dege1}
\end{equation}%
and
\begin{equation}
\left( \partial _{r}\partial _{r^{\prime }}+\frac{q^{2}m^{2}}{rr^{\prime }}%
\right) \int_{0}^{u}\frac{dx}{x}\,g(r,r^{\prime },x)=4ue^{-bw}\partial
_{w}I_{qm}(w),  \label{dege2}
\end{equation}%
with the notations (\ref{wbn}). In addition we have%
\begin{equation}
\left( \partial _{r}\partial _{r^{\prime }}+\frac{q^{2}m^{2}}{rr^{\prime }}%
\right) g(r,r^{\prime },u)=4u\partial _{w}\left( we^{-bw}\partial
_{w}I_{qm}(w)\right) .  \label{geder}
\end{equation}

Hence, we get the following results%
\begin{eqnarray}
\mathcal{J}_{\nu }^{(n,p)} &=&(4\pi )^{D/2-1}\int_{0}^{\infty }du\,u^{D/2-2}%
\left[ 4u\left( D/2-1-|\Delta \mathbf{z}|^{2}u\right) \right] ^{n}e^{u\left(
\eta ^{2}+\eta ^{\prime 2}-|\Delta \mathbf{z}|^{2}\right) }  \notag \\
&&\times K_{\nu }(2\eta \eta ^{\prime }u)\left( 4u^{2}\partial _{w}\right)
^{p-1}\left[ we^{-bw}I_{qm}(w)\right] ,  \label{Jnp3}
\end{eqnarray}%
for $p=1,2$ and%
\begin{eqnarray}
\mathcal{J}_{\nu }^{(1,0)} &=&\frac{1}{4}(4\pi )^{D/2-1}\int_{0}^{\infty
}du\,u^{D/2-2}\left[ 4u\left( D/2-1-|\Delta \mathbf{z}|^{2}u\right) \right]
^{n}e^{u\left( \eta ^{2}+\eta ^{\prime 2}-|\Delta \mathbf{z}|^{2}\right) }
\notag \\
&&\times K_{\nu }(2\eta \eta ^{\prime }u)\int_{0}^{u}\frac{dx}{x}%
\,g(r,r^{\prime },x).  \label{J10}
\end{eqnarray}%
The results for the latter integral after the action of the operators,
appearing in (\ref{CE1}), read%
\begin{eqnarray}
\left( r\partial _{r}+r^{\prime }\partial _{r^{\prime }}\right) \mathcal{J}%
_{\nu }^{(1,0)} &=&\frac{1}{2}(4\pi )^{D/2-1}\int_{0}^{\infty }du\,u^{D/2-2}%
\left[ 4u\left( D/2-1-|\Delta \mathbf{z}|^{2}u\right) \right] ^{n}  \notag \\
&&\times e^{u\left( \eta ^{2}+\eta ^{\prime 2}-|\Delta \mathbf{z}%
|^{2}\right) }K_{\nu }(2\eta \eta ^{\prime }u)e^{-bw}I_{qm}(w),  \label{J10a}
\end{eqnarray}%
and%
\begin{eqnarray}
\left( \partial _{r}\partial _{r^{\prime }}+\frac{q^{2}m^{2}}{rr^{\prime }}%
\right) \mathcal{J}_{\nu }^{(1,0)} &=&(4\pi )^{D/2-1}\int_{0}^{\infty
}du\,u^{D/2-1}\left[ 4u\left( D/2-1-|\Delta \mathbf{z}|^{2}u\right) \right]
^{n}  \notag \\
&&\times e^{u\left( \eta ^{2}+\eta ^{\prime 2}-|\Delta \mathbf{z}%
|^{2}\right) }K_{\nu }(2\eta \eta ^{\prime }u)\partial _{w}\left[
we^{-bw}\partial _{w}I_{qm}(w)\right] .  \label{J10b}
\end{eqnarray}%
with $w$ and $b$ defined by (\ref{wbn}).

\end{document}